\documentclass[twocolumn,pra,amsmath,amssymb,showpacs,floatfix]{revtex4}

\usepackage{epsfig,amstext,amssymb,amsmath}

\begin{document}

\title{Atomic Josephson vortex}
\author{V.M. Kaurov and A.B. Kuklov}
\affiliation
{Department of Engineering Science and Physics,
The College of Staten Island, CUNY, 
Staten Island, New York 10314} 
\date{\today}

\begin{abstract}
We show that Josephson vortices in a quasi-1D atomic Bose Josephson junction can be controllably 
manipulated by imposing a difference of chemical potentials on
the atomic BEC waveguides forming the junction. This effect, which has its origin in the Berry phase structure of a vortex,
turns out to be very robust in the whole range of the parameters where such vortices can exist.
We also propose that a Josephson vortex can be created by the phase imprinting technique and can be 
identified by a specific {\it tangential} feature in the interference picture produced by
expanding clouds released from the waveguides. 

\end{abstract}


\maketitle

\section{Introduction}
From the very beginning of the experimental achievement 
of Bose-Einstein condensation in trapped ultracold gases, 
various solitonic and topological collective
structures, such as dark \cite{DS} and bright \cite{brsl}
 solitons, vortices \cite{vortex}, skyrmions \cite{skyr}, etc., became 
objects of great theoretical and experimental interest. These configurations 
are direct manifestations of macroscopic coherence of BEC and many 
of them were repeatedly created and detected. 

In our previous work \cite{orig} it has been shown that yet another stable object, 
namely atomic Bose Josephson vortex (JV), can exist in a long Bose Josephson junction (BJJ).
Such junction can be formed between two parallel quasi-1D Bose Einstein condensates 
coupled by tunneling. The circulating atomic 
supercurrents --- counter-propagating in each waveguide and closing the loop
between them ---  represent a neutral analog of the Josephson vortex studied in 
superconducting Josephson junctions in great detail \cite{barone}. While bearing 
similarities with the charged case, the absence of the Meissner effect makes 
the local formulation of the BJJ solution possible only in a quasi-1D geometry. 
The most interesting feature of the Bose JV is its interconversion 
into a dark soliton (DS) \cite{orig}.  Accordingly, the description of the Bose JV necessarily 
involves both the phase and the amplitude of the
wavefunction. This instability, which can be controlled by tuning the Josephson coupling, may potentially
be utilized in several ways \cite{orig}. Here we show that the Berry phase
term is responsible for a force produced on the JV by a time-dependent difference of the
chemical potentials $\delta \mu$ applied between the waveguides. Once the JV speed as a whole reaches
a certain critical value, the instability destroying the supercurrent circulation develops so that
the JV transforms into a moving DS (grey soliton) which is essentially insensitive
to $\delta \mu$. We also 
discuss how the phase imprinting method and the interference after some free expansion 
can be employed for creating and detecting the JV. As it turns out, the formation of the relative phase between
the waveguides is quite insensitive to the characteristics of the imprinting beams once certain topological
requirements are satisfied. The JV introduces a specific {\it tangential} feature into the interference pattern,
which should allow an unambigious identification of the JV as well as of the interconversion effect \cite{orig}.

Experimental tools with great potential for creation, manipulation and detection of the JV have already become available and continue to develop. Long BJJ consisting of two parallel coupled atomic waveguides was experimentally realized with two different and equally versatile techniques capable of producing double-well potential of necessary geometry. First technique uses counterpropagating laser beams to form an optical dipole trap \cite{Kett2} and second one uses a microfabricated atom chip with purely magnetic potential \cite{mgfch}. In both cases relative coherence of BECs has been revealed in the interference fringes.

\section{Atomic quasi-1d Josephson vortex}

Below we will introduce a phenomenological description of the atomic
JV which captures three main effects: i) The JV instability with respect
to its interconversion into the DS as the Josephson coupling strength exceeds
certain threshold \cite{orig}; ii) The critical speed effect;
iii) The effect of force on the JV produced by uniform $\delta \dot{\mu}$. We will also
derive the corresponding phenomenological coefficients in terms of the microscopic
model \cite{orig} close to the threshold and will consider the opposite limit --- very small
couplings, when the JV can be well described by the Sine-Gordon (SG) equation.

\subsection{Phenomenological free energy}

Phenomenological description helps elucidating generic features which 
do not depend qualitatively on a particular microscopic model.
The main characteristic of the JV is a presence of a localized supercurrent
circulation degenerate with respect to its orientation.
This circulation can be decsribed by persistent currents $J_{1,2}(x,t)$ along the first and
the second waveguides, respectively, as well as by tunneling Josephson currents between
the waveguides. Here $x$ is a spatial coordinate along two parallel waveguides and $t$ stands
for time. If the JV is stationary, the currents $J_{1,2}(x,t)$  are localized within some typical
distance and flow in opposite
directions, and, therefore, the difference $J(t)=\int dx (J_1(x,t)- J_2(x,t))$ is a good
global measure of the circulation. Obviously, the quantity $P(t)=\int dx (J_1(x,t)+ J_2(x,t))$
is linear momentum of the JV. It is zero, if the JV is stationary, and it must acquire
a finite value, if it is moving as a whole. The total momentum is assigned to center of mass
of the JV positioned somewhere at $X_0(t)$ along the junction. 
This generic picture
is well known from the phenomenological description of the JV in superconductors \cite{barone},
 where the SG model is sufficient to find the relation between $P(t)$
and the center of mass velocity $V=\dot{X}_0$ through the JV effective mass $M$. 

It is important that in the SG model the circulation is essentially a topologically conserved
quantity (provided the contribution to the circulation from the tunneling currents are negligible). 
Thus, this model 
does not contain the intercoversion effect \cite{orig} taking place in a quasi-1D geometry.
Incorporating density fluctuations shows that the center
of mass position $X_0$ can be associated with a dip in the condensate densities in each
waveguide. This dip becomes more pronounced as the Josephson coupling $\gamma$ strengthens, and, finally,
it becomes zero of the densities at some critical value $\gamma =\gamma_c$, 
which signals a moment of the transformation of the JV into the DS. At this
point and for higher $\gamma$ no more current circulation $J$ exists. Obviously, this
effect can be interpreted in terms of spontaneous breaking of time reversal symmetry 
for $\gamma < \gamma_c$ with the order parameter $J \neq 0$. Since $J$ is just one
degree of freedom the corresponding "phase transition" occurs in zero spatial
dimensions and therefore cannot be considered as a true phase transition. Yet, the
corresponding free energy of the system containing just one DS or JV can be represented
in a sense of the Landau expansion of the effective action
close to the "critical" point $\gamma \approx \gamma_c$  as
\begin{equation}\label{lagr_ef}
{\cal L}_{eff} = \dot{X}_0P + \dot{Q}J -{\cal H}_{eff},
\end{equation}
with the effective Hamiltonian being
\begin{equation}\label{haml_ef}
{\cal H}=\frac{P^2}{2M_+} + \frac{Q^2}{2M_-} + \alpha (\gamma - \gamma_c) J^2
+ c_1 J^4 - c_2 \delta \mu P J+ c_3 J^2P^2. 
\end{equation}
where $Q(t)$ stands for a variable canonically conjugate $J$;      
$M_+,\, M_-,\, \alpha >0,\, c_1>0,\, c_2,\, c_3 >0$ are phenomenological
coefficients dependent on a particular form of a microscopic description.
These coefficients will be determined later within the variational approach applied
to the model \cite{orig}. 

The Lagrangian (\ref{lagr_ef},\ref{haml_ef}) describes the effect i), that is,
spontaneous formation
of the circulation $J\neq 0$ for $\gamma < \gamma_c$ \cite{orig} ($c_1>0$ insures stability
beyond linear approximation) as well as coupling
between the center of mass motion and the internal circulation. 

The term $\sim c_3$
accounts for the effect ii): as $P \sim V$ exceeds some critical value,
the effective coefficient $\alpha (\gamma - \gamma_c) + c_3 P^2$ becomes positive, which
restores the time-reversal symmetry so that the circulation $J=0$. 

The term $\sim c_2$ is responsible for the effect iii), that is, for
the force induced by $\delta \mu$. This term is symmetric 
with respect to exchanging of the waveguides and conforms with the time-reversal
symmetry. Indeed, swapping the waveguides positions changes $\delta \mu \to -\delta \mu$ and
 $J \to -J$. The time-reversal leaves the product $JP$ invariant.
The term $ \delta \mu P J$ is responsible for the force $\sim \delta \dot{\mu} J$ on the JV.
Its origin can be understood as follows:
When $\delta \mu \neq 0$, a number of atoms $\sim \delta \mu $ starts tunneling between the waveguides.
 Accordingly, the first and the second waveguides acquire the total 
linear momenta $\sim \delta \mu \int dx J_1$ and $\sim - \delta \mu \int dx J_2$, respectively.
Hence, the total momentum $P$ attains a nonzero value $\sim \delta \mu J$
as long as the circulation $J$ is finite. The force, then, is $\dot{P} \sim \delta \dot{\mu} J$.
It can also be viewed as a force due to the bias introduced by the tunneling currents 
driven by $\delta \dot{\mu}$ (compare with the bias term in the SG equation
\cite{barone}).

In the presence of dissipation one should introduce the dissipative function in terms
of the velocities $\dot{X}_0$ and $\dot{Q}$ as
\begin{equation}\label{dis_ef}
{\cal F}=\frac{\dot{X}_0^2}{2\sigma_+} + \frac{\dot{Q}^2}{2\sigma_-},  
\end{equation}
with some kinetic coefficients $\sigma_{\pm} >0$.
Then the standard variational procedure with respect to the conjugate
variables yields the equations of motion
\begin{eqnarray}
V-\left(\frac{1}{M_+} + 2c_3J^2\right)P - c_2\delta \mu J=0,
\label{dP}\\
\dot{P}+\frac{V}{\sigma_+}=0,
\label{dV}\\
\dot{Q}-2\left(\alpha (\gamma - \gamma_c) + c_3P^2\right)J - 4c_1J^3 -c_2\delta \mu P=0,
\label{dJ}\\
\dot{J}+ \frac{Q}{M_-} + \frac{\dot{Q}}{\sigma_-}=0.
\label{dQ}
\end{eqnarray}
This system describes a quite complex non-linear dynamics of 
the circulation $J$ coupled to the center of mass motion.
Besides small oscillations of $J$ around equilibrium, strongly
non-linear evolution of $J $ corresponding to the phase-slip,
during which $J$ changes sign, can occur as $\gamma$ approaches $\gamma_c$
from below. The term $\sim c_3$ in eq.(\ref{dJ}) is responsible 
for strong coupling of the phase-slip to the motion of the JV as a whole.
Furthermore, the terms $\sim \delta \mu$ introduces the parametrical driving.
Thus, all three effects mentioned above cannot in general be well
separated from each other, and general solution of the
system can only be obtained numerically. However, considering certain limiting cases
is helpfull for making these effects independent from
each other. So, in what follows we will consider such limiting situations.
In reality, of course, the above system should be analyzed numerically.

Let us, first, consider the case $\delta \mu=0$ and $P=0$, that is,
stationary JV.
Then, eqs.(\ref{dJ},\ref{dQ}) are essentially those discussed in ref.\cite{orig}
which describe the dynamics of the JV-DS interconversion which involves
the phase-slip. Here we will not discuss
this effect (that is, i)) any further. 

The effect ii) makes sense to consider
only in the case of small dissipation with respect the center of mass motion.
So, we set $1/\sigma_+ =0$ and, keeping $\delta \mu=0$, find
that the generalized momentum is conserved $P=const$. Then, 
eq.(\ref{dP}) gives $P=M_+ V$ in the limit of small $J$. 
Further substitution into eq.(\ref{dJ}) yields 
$-2\left(\alpha (\gamma - \gamma_c) + c_3M_+^2V^2\right)J - 4c_1J^3=0$
in equilibrium. A non-trivial solution exists only when
the coefficient $\alpha (\gamma - \gamma_c) + c_3M_+^2V^2$ in front the
linear term in eq.(\ref{dJ}) is negative.  Accordingly, the critical 
velocity becomes
\begin{equation}\label{V_cr}
V_c=V_1\sqrt{\gamma_c - \gamma} \ ,\quad V_1  = \sqrt {\frac{\alpha }{(c_3 M_ + ^2 )}},
\end{equation}
above which no quasi-static solution for the circulation $J\neq 0$ exists.

Now, let's consider the effect iii) where $\delta \mu \neq 0$ and small. Then, we
 assume that velocities are much smaller then the critical one (\ref{V_cr})
and the JV circulation parameter $J$ is in equilibrium close
to its value $J_0= \pm \sqrt{\alpha(\gamma_c -\gamma)/2c_1}$ determined when $V=0$.
In this situation, one should ignore the term $\sim c_2$ in eq.(\ref{dJ}) because
it describes just a small correction to $J_0$.
Subsequently, $J$ can be replaced by $J_0$ in eq.(\ref{dP}) and term $\sim c_3$
should be dropped because it describes a small correction $\sim \gamma_c -\gamma$
to the finite term $1/M_+$.    
Then, excluding $P$, one finds
that the center of mass velocity obeys the equation
\begin{equation}\label{V_eq}
\dot{V}+ \frac{V}{M_+\sigma_+}=\frac{f(t)}{M_+},\quad f(t)= c_2M_+ J\dot{\delta \mu}.
\end{equation}
As discussed above, the term
$f(t)$ describes the force induced by time dependence of the difference of the chemical
potentials $\delta \mu$. The relation of this force to the Berry phase term in the full "microscopic" action
will be considered below.  
 
\subsection{Variational approach}
It is worth noting that all the phenomenological
coefficients can be derived from a "microscopic" Lagrangian and a 
dissipative function. Here we will use  a simplified approach which
ignores dissipation and will consider model \cite{orig} as the "microscopic" Lagrangian.
In terms of the fields $\psi_{1,2}$
describing each waveguide the Lagrangian, 
\begin{equation}\label{lagr}
{\cal L} = {\cal L}_{B} - {\cal H},
\end{equation}
is given by the 
Berry term
\begin{equation}\label{berry}
 {\cal L}_{B}={\rm Re} \int dx\left[ {i\hbar (\psi _1^* \dot \psi _1  + \psi _2^* \dot \psi _2 )} \right] ,
\end{equation}
and by the
Hamiltonian
\begin{equation}\label{haml}
{\cal H}={\cal H}_1+{\cal H}_2+{\cal H}_{12},
\end{equation}
consisting of the each waveguide terms 
\begin{equation}\label{H_k}
{\cal H}_k=\int dx[\frac{\hbar^2}{2m}|\nabla \psi_k|^2 +\frac{g}{2}|\psi_k|^4 - \mu_k|\psi_k|^2],
\end{equation}
with $k=1,2$, and of the contribution which describes the Josephson tunneling between the waveguides
\begin{equation}\label{H_12}
{\cal H}_{12}  =  - \int dx\gamma (\psi _1^* \psi _2  + \psi _2^* \psi _1 ),
\end{equation}
where $\mu_k$ are waveguide's chemical potentials and the integration $\int dx ...$
is performed along the waveguides.

The equations of motion following from eqs.(\ref{lagr}-\ref{H_12}) 
in the units $\hbar =1, m=1$, with the unit of length given by the
healing length ~$l_c=1/\sqrt{\mu}$ for $\mu = (\mu_1 + \mu_2)/2$ 
and the unit of time determined by ~$t_0=1/\mu$, are
\begin{eqnarray}
i\dot{\psi}_1 &=&
-\frac{\nabla^2}{2} \psi_1 - (1+ \eta)\psi_1 
+|\psi_1|^2\psi_1 - \nu \psi_2;
\label{eq_1}\\
i\dot{\psi}_2
&=&-\frac{\nabla^2}{2} \psi_2 - (1-\eta)\psi_2 
+|\psi_2|^2\psi_2 - \nu \psi_1.
\label{eq_2}
\end{eqnarray}
Here the wave functions were 
transformed as ~$\psi_k \to \sqrt{n_0}\psi_k $, where 
$n_0=\mu/g$ is the average 1D density of a single uncoupled 
($\gamma=0$) waveguide; the quantity $\nu$ represents the dimensionless Josephson
coupling $\nu =\gamma /\mu$, and $\eta$ stands for the difference of chemical potentials
rescaled by $2\mu$. 
These equations admit exact JV stationary solution discussed in detail in ref.\cite{orig}.
For small velocity of the JV, it is natural to use
the variational ansatz which coincides with the stationary solution
at $V=0$. Thus, we choose
\begin{equation}\label{anz2}
\Psi _{1,2}  = \sqrt {n_{1,2} }\,\,\,{\operatorname{th} 
\left( s \left( {x - X_0(t)} \right) \right) +\frac{i\sqrt{s}Q_{1,2}}
{{\operatorname{ch}\left( s \left( {x - X_0(t)} \right) \right)}}} 
\end{equation}
with the parameter $s$ giving the JV size.
The bulk densities $n_{1,2}$ can be obtained in the thermodynamical limit  (when
no JV is present) from eqs.(\ref{eq_1},\ref{eq_2}).
Since we are interested in small deviations only, the corresponding explicit 
expressions are 
\begin{equation}\label{n12}
n_{1,2} = (1+\nu)\left(1 \pm \frac{\eta}{1+ 2\nu} + o(\eta^2)\right), 
\end{equation}
where the indexes $1,2$ correspond to $\pm$, respectively.

The stationary solution \cite{orig} can be obtained from the ansatz (\ref{anz2})
by setting $X_0=const$,
$\sqrt{s}Q_1=-\sqrt{s}Q_2=\pm \sqrt{1-3\nu}$ and $s=2\sqrt{\nu}$ (compare with ref.\cite{orig}). 
Considering complex $Q_{1,2}$ and real $X_0$ as slow dynamical variables,
one can substitute the ansatz (\ref{anz2}) into the Lagrangian   
(\ref{lagr}-\ref{H_12}) and perform explicit integration over $x $ .
This procedure generates the effective Lagrangian ${\cal L}_e$ in terms of the
variables $ Q_{1,2}, \, X_0, \, s$ and their time-derivatives.
Obviously, such procedure is in line with separation of fast and slow
variables, so that only slow dynamics should be considered to full
extent. Close to the interconversion instability ($\nu \approx \nu_c=\gamma_c/\mu $), the slow variables are
$X_0$ and $J$ .  
The variable $s$ describes fast adjustment of the JV size.
It is important that it is not dynamical within the chosen ansatz. Indeed, as can be seen,
  the effective Lagrangian
does not contain $\dot{s}$. The choice of $s$ is dictated by conservation of total
number of particles during dynamical evolution of the other parameters. Calculating
the depletion $\delta N$ of the number of particles caused by the presence of the JV, we find
\begin{equation}\label{depl}
\delta N=-\frac{2(n_1 + n_2)}{s} + 2(|Q_1|^2 + |Q_2|^2) =C_N,
\end{equation}
where the constant $C_N$ is determined for the stationary JV by setting
all the time derivatives to zero and minimizing the effective energy with
respect to $Q_{1,2}$ and $s$. Considering small values  $\eta$, it is enough to
set $n_1 + n_2=2(1+\nu)$, which is the equilibrium value. Here we will consider
 values $Q_{1,2} \to 0$, so that the explicit solution of eq.(\ref{depl}) for $s$ becomes
\begin{equation}\label{s}
s=\sqrt{1+\nu} + \frac{1-3\nu}{2\sqrt{1+\nu}}
 - \frac{|Q_1|^2 +|Q_2|^2}{2}.
\end{equation}
As discussed in ref.\cite{orig}, the value $\nu=\nu_c=1/3$ is the critical point
below which the JV forms spontaneously from the DS. Thus, the smallness of
$Q_{1,2}$ automatically implies a proximity to the critical
point. Then, for consistency of the effective action expressed in powers of $Q_{1,2}$, the value $\nu$ should be set to
$\nu_c$ except in the quadratic term vanishing at the critical point. 

It is worth discussing, first, the structure of the Berry-term part (\ref{berry}) 
of the full action. 
As mentioned above, the cross term $\sim \eta P J$ leading to the
force on the JV $\sim \dot{\eta}$ in the Lagrangian (\ref{lagr_ef})
can be viewed as generated by the Berry phase effect. Indeed, 
the Berry part is ${\cal L}_B=\int dx( -\rho_1 \dot{\varphi}_1 -\rho_2 \dot{\varphi}_2)$, where
$\rho_{k}$ and $\varphi_{k}$ are density and the phase, respectively, in the $k$-th waveguide.
In the static solution \cite{orig} as well as in the ansatz (\ref{anz2}) each phase
changes by $\pm \pi$, so that,e.g., if at $x=-\infty$ one finds $\varphi_1=\varphi_2=0$, then,
at $x=+\infty $  there is $\varphi_1-\varphi_2=2\pi$. If the JV is moving slowly, then
$ \dot{\varphi}_k\approx \dot{X}_0 \nabla \varphi_k$. Thus, a substitution into the Berry part
gives  
${\cal L}_B\approx \dot{X}_0\int dx (\rho_1\nabla \varphi_1 + \rho_2\nabla\varphi_2)\approx \pi \dot{X}_0 (\rho_1 - \rho_2)$,
where spatial variations of the densities are ignored. Flipping the time derivative and realizing
that $ \rho_1 - \rho_2 \sim \eta$, one finds ${\cal L}_B\sim - X_0 \dot{\eta}$, which is the work
done while making a displacement $X_0$ by the force $\sim \dot{\eta}$.

The relation between the observables $P,\, J$ can be obtained as a result of substituting
the ansatz (\ref{anz2}) into (\ref{berry}). This yields
${\cal L}_B=\dot X_0 P - 2(\dot B_ +  Q_ +   + \dot B_ -  Q_ -  )$, where $Q_+={\rm Re}(Q_1 + Q_2)$,
$ Q_-={\rm Re} (Q_1 - Q_2)$, $ B_+={\rm Im} (Q_1 + Q_2) $ and $ B_-={\rm Im} (Q_1 - Q_2) $.
The total momentum
$P=-i\int dx(\psi^*_1\nabla \psi_1 +\psi^*_2\nabla \psi_2)$ and the
supercurrent circulation $J=-i\int dx(\psi^*_1\nabla \psi_1 -\psi^*_2\nabla \psi_2)$
are given as
\begin{eqnarray}
P=-\pi (1+\nu)^{3/4}\left[Q_+ + \frac{\eta Q_-}{2(1+2\nu)}\right],
\label{P}\\
J=-\pi (1+\nu)^{3/4}\left[Q_- + \frac{\eta Q_+}{2(1+2\nu)}\right].
\label{J}
\end{eqnarray}
Thus, the parameters $Q_{\pm}$ can uniquely be expressed in terms of $P,\,J$.
In particular, $Q_+ \sim P$ for $\eta=0$.
The Berry phase effect discussed above is reflected in the part $\sim \eta$ of eq.(\ref{P}).
It is also clear that the variable conjugated to $J$ in eq.(\ref{lagr_ef}) is $Q\sim B_-$.
Further analysis shows that
the quantity $ B_+={\rm Im} (Q_1 + Q_2) $
enters ${\cal L}_B$ in a combination $\sim \dot{B}_+ P - r_{b} B_+^2 + o(B^4_+)$
with some $r_{b}>0$. Thus, $B_+$ would generate higher time
derivatives with respect to $X_0$, which should be neglected as long as the JV motion
is slow. Accordingly, it is reasonable to set $B_+=0$ in eq.(\ref{anz2}), so that
$Q_{1,2}$ are chosen in the form 
\begin{equation}\label{anz3}
Q_{1,2} = \frac{Q_+ \pm Q_-}{2} \pm i\frac{B_-}{2},
\end{equation}
with the indexes $1,2$ corresponding to $\pm$, respectively.
Finally, employing the ansatz (\ref{anz2},\ref{anz3}) in the
"microscopic" Lagrangian (\ref{lagr}-\ref{H_12}) and expressing the
variables (\ref{anz3}) in terms of the observables $P,\, X_0, \, Q, \,J$ as described above
one arrives at the effective Lagrangian in the form (\ref{lagr_ef}), where the
coefficients (in the chosen units) are
\begin{eqnarray}
\frac{1}{M_+}&=& - \frac{{\sqrt 3 }}{{2\pi ^2 }},
\label{mass_+}\\ 
\frac{1}{M_-}&=&\frac{{32\pi ^2 }}{{27\sqrt 3 }},
\label{mass_-} \\
\alpha \left({\gamma -\gamma _c }\right)&=&\frac{{3\sqrt 3 }}{{4\pi ^2 }}\left( {\nu  - \frac{1}{3}} \right) ,
\label{alpha}\\
 c_1&=& \frac{{3\sqrt 3 }}{{128\pi ^4 }},  
\label{c_1}\\
c_2&=&  \frac{{2\sqrt 3 }}{{5\pi ^2 }}, 
\label{c_2}\\
c_3&=& \frac{{9\sqrt 3 }}{{64\pi ^4 }}.
\label{c_3}
\end{eqnarray}
In these expressions, the difference $\sim \nu -\nu_c$
has been ignored except in the quadratic coefficient (\ref{alpha}),
which determines the instability. 

It is worth noting that far from the instability
(that is, $\nu \ll \nu_c$) no simple expansion for the effective Lagrangian in terms of 
powers of $Q_{1,2}$ can be obtained.
However, it is clear that in this limit, the variations of the density
can be ignored as long as the JV is close to its internal equilibrium.
Thus, the effective action can simply be obtained by considering
fluctuations of the densities being small.
This will lead to the SG model. For completness, we will
consider this limit below.

\subsection{Sine-Gordon approximation}
Setting
$\psi_1=\sqrt{n_1 +n'_1(x,t)}\exp(i\varphi),\,\, 
\psi_2=\sqrt{n_2 +n'_2(x,t)}\exp(i\varphi)$, with
$ n_{1,2}$ being uniform background densities given by
eq.(\ref{n12}) and $n'_{1,2}$ describing small fluctuations,
and substituting this
into the action (\ref{lagr}) one obtains the effective Lagrangian
in terms of the relative phase $\varphi$ only as
\begin{eqnarray}
{\cal L}_{SG}&=& \int dx [ (n_2 -n_1) \dot{\varphi}
+ \dot{\varphi}^2  
\nonumber \\
&-&(1+ \nu) (\nabla \varphi)^2
+ 2\nu(1+\nu)\cos(2\varphi)],
\label{SG}
\end{eqnarray}
where  $n'_{1,2}$ were eliminated
in the long wave limit, with the gradients of $n'_{1,2}$ ignored;
the difference $n_2 - n_1 \neq 0$ has only been retained in the
Berry term.

The variation yields
\begin{eqnarray}
\ddot{\varphi} - (1+\nu)\nabla^2\varphi + 4\nu(1+\nu)\sin(2\varphi)=\dot{n}_1 -\dot{n}_2
\label{SGE}
\end{eqnarray}
the biased SG equation with its well known solution
in the static limit (see in \cite{barone}). It is important
to note that this equation exhibits Lorentz invariance (for
$\dot{n}_1 -\dot{n}_2=0$). Hence, no the critical velocity
effect ii) can be described in the SG approximation.

Here we will be interested the limit of slow velocity $V$ and 
small $\dot{n}_1 -\dot{n}_2$
in eq.(\ref{SGE}). Then, 
one can use the static SG solution, with its center of mass
velocity $V$ being a slow variable \cite{barone}, in the Lagrangian (\ref{SG}).
Then, integrating explicitly, one finds 
\begin{eqnarray}
{\cal L}_{SG}&=& \int dx [ -2\pi \eta V
+ \frac{M_{SG} V^2}{2}]  
\label{SG_X}
\end{eqnarray}
where the effective mass is
\begin{eqnarray}
M_{SG}=4\sqrt{2\nu}. 
\label{M_SG}
\end{eqnarray}

Varying with respect
to $X_0$ and including the dissipative function, one
arrives at the equation similar to eq.(\ref{V_eq}):
\begin{eqnarray}
\dot{V}+\frac{V}{\sigma_+ M_{SG}}= \frac{2\pi \dot{\eta}}{M_{SG}},
\label{V_SG}
\end{eqnarray}
which describes the effect of force induced by the Berry term.
Thus, while a particular expression for the force
depends on the proximity to $\nu_c$, the effect persists for all
values $\nu <\nu_c$.


\subsection{Numerical simulation of the Berry phase effect}
\begin{figure}[h]
\begin{center}
\epsfxsize=8.5cm
\epsfbox{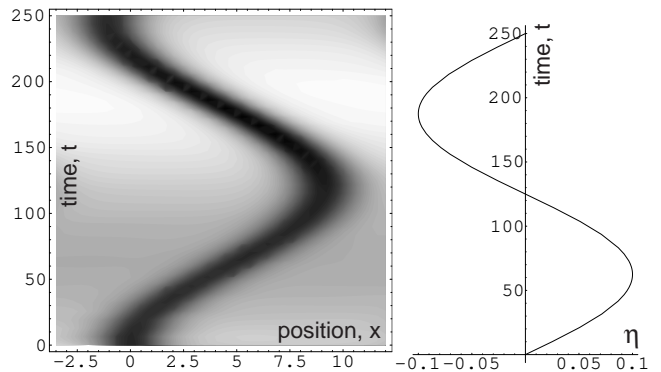}
\caption{
\label{move}
Motion of the JV along the junction (left) generated by change of relative 
chemical potential $\eta$ (right) from numerical simulations of the system 
(\ref{eq_1}, \ref{eq_2}). Left figure: the atomic density in a single waveguide is represented by intensity of white color; the JV path in $x$-$t$ coordinates corresponds to the dark color (the density depletion in the JV center) relative to the white color of the background. Here $\nu=0.1$, and the dissipative term from ref.\cite{orig}
is given by the value of the kinetic parameter $\tilde \sigma =2$. The units of x and t 
are the coherence length $l_c$ and the coherence time $t_0$, respectively.
} 
\end{center}
\end{figure}

Sensitivity to the difference of the chemical potentials $\eta$ can be a useful tool for manipulation of the JV position 
in the junction. To demonstrate this numerically we used $\eta$ as an externally controlled variable to displace the JV 
on a distance much greater than its size and, then, to return it to 
its original position. The result of the simulations of the full system
(\ref{eq_1}, \ref{eq_2}) is displayed on 
FIG.\ref{move}. The plot on the left represents 
density of a single waveguide by the intensity of white color. The dark 
curve is  a trajectory of the JV center, where its density is 
minimal. Shown on the right, the time-dependence $\eta$ is chosen as
$\eta (t) = 0.1\sin(\omega t)$, with 
$\omega=\pi/125$. 
The simulations have been performed with the dissipative term
introduced in ref.\cite{orig}.
It is important to note that the regime chosen for numerical
simulations is neither close to the critical point nor to
the SG limit. Yet, the time dependence $X_0(t)= A(1-\cos(\omega t))$,
which follows from eqs.(\ref{V_eq},\ref{V_SG}) in the limit of small
damping and where A is some
amplitude, is consistent with the simulations with the reservation
that the numerical value of $A$ is not reproduced correctly
by either limiting approximation: while eq.(\ref{V_eq}) underestimates $A$ by a factor of 2, the
SG limit (\ref{V_SG}) overestimates it by a factor of 4.

\section{ Creation by phase imprinting}

The JV can be formed as a result of the
decay of the DS, once the Josephson coupling $\gamma$ is reduced below a critical value $\gamma_c$ (or, in 
the chosen units, $\nu$, with the critical value being $\nu_c=1/3$) \cite{orig}. 
An alternative method is the phase imprinting. It is already a well established  experimental tool for creation of the DS \cite{DS}. It consists of exposing a BEC to a pulse of a far detuned laser beam which acts as a temporary external potential $U(x,t)$. According to the impulse approximation, the duration of the pulse $\delta t$ must be short compared to
the correlation time of the condensate $t_0 = 1/\mu$ , so that no change of the BEC density occurs during the
pulse --- atoms just acquire finite speeds without performing any significant displacements.
\begin{figure}[h]
\begin{center}
\epsfxsize=8.5cm
\epsfbox{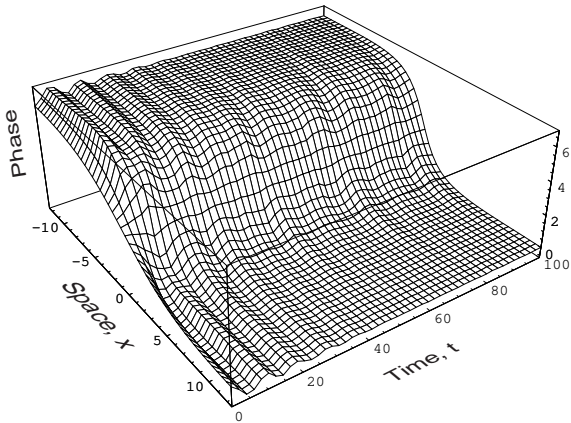}
\caption{
\label{phimp}
Evolution of the relative phase after the imprinting as follows from eqs.(\ref{eq_1},\ref{eq_2}). Small dissipative
term discussed in ref.\cite{orig} with $\tilde \sigma =6$ has been added; $\nu=0.1$. The units of x and t are the coherence length $l_c$ and the coherence time $t_0$ respectively.}
\end{center}
\end{figure}
 The phase, on the other hand, accumulates according to 
$\delta\varphi (x) = \int dt U(x,t)$ and the wave function $\psi$ before the pulse is transformed 
as $\psi  \to e^{ - i\delta\varphi } \psi$ after the pulse. To create a DS, one needs to 
expose a half-plane of an elongated BEC to a laser pulse with a spatial variation 
reminiscent of the typical DS phase profile (the $\pi$-step). 
In order to produce the JV, one needs to apply a pulse with spatial profiles $U_{1,2}(x,t)$ specific for
each waveguide. These should reflect the structure of the JV phases in each waveguide: $\varphi_{1,2}=0$
at ,e.g., $x=-\infty$ and $\varphi_1=-\varphi_2= \pm \pi$ at $x= +\infty$, with
smooth transition in between at a typical length comparable to
the JV size ~$= 1/(2\sqrt{\nu})$. Accordingly, $U_{1,2}(x,t)=0 $ at $x=-\infty$ and
$U_1(x,t)=-U_2(x,t)$ at $x= +\infty$, with the "crossover" region being approximately equal
to the JV size and the time integral $\int dt U_1(x,t)=\pi$ at $x \to +\infty$.

It is important to note that, once the above "topological" requirements are satisfied,
the JV forms with minimal disturbances regardless of other details of the laser beams profiles.
The most robust characteristic of the evolving solution turns out to be the phase difference between the waveguides.
 In the case of the DS, the complete density depletion must form at the DS center. Thus, the adjustment is accompanied by a strong perturbation in the form of the density waves \cite{DS}. 
The depletion at the JV center is also strong, if $\nu$ is close to the critical value
$1/3$. Accordingly, the densities in each waveguide will experience significant perturbations. 
Yet, the phase difference relaxes quite smoothly to the equilibrium profile.
To demonstrate this feature, we ran numerical simulations with the initially imprinted profile
of the phases given by the $\tanh$-type variations of the light intensities as described above.
The result of the following evolution of the phase difference is represented on FIG.\ref{phimp}.
As can be seen, the equilibrium phase profile establishes after few relatively small
oscillations even though the initial extension of the (imprinted) phase was about
two times longer than the equilibrium JV size.

\section{ Detection by interference}
Experimental visualization is typically done by absorption imaging \cite{Kett1,Kett2,Kett4},
with its intensity proportional to the density $n$ of the expanding cloud.
In contrast to bulk vortices, which can be detected by observing
their cores, the JV does not have a core. Yet, the phases exhibit the
$\pi$-jumps. Thus, the interference of the expanding clouds released
from the waveguides should demonstrate a corresponding feature.

In quasi-1D regime a good approximation for the waveguides wave functions
in transverse directions 
is the Gaussians $G(y,z)= \exp \left( - (y^2+z^2) /2d^2  \right)$, with
$d$ being a typical width of each waveguide.
Thus, in 3D, the two waveguides 
separated by a distance $2z_0$ can be described by the following ansatz:
\begin{eqnarray}
\Psi _0 (\vec R) = \Psi _0^ +   + \Psi _0^-
\label{int_1}\\
\Psi _0^ \pm   =f(x) \psi _{1,2} (x)G(y,z \pm z_0)
\label{int_2}
\end{eqnarray}
where $\psi_{1,2}(x)$ are the solutions (\ref{anz2}) corresponding to either 
the DS ($Q_{1,2}=0$) or to the JV and the sign $\pm$ is different for different waveguides.
The envelope $f(x) = \left( {1 - (4x^2 /L)} \right)$, with $L$, 
the axial system size, being much
larger than the JV, reflects finiteness of the BEC clouds.

As long as the transverse dimension $d$ is much smaller than any 
axial feature, the expansion occurs primarily in the transverse
direction. Thus, the density decreases, practically  instantaneously, so that the expansion is essentially free of
interaction. 
This can be formulated as the requirement $(d/l_c)^2 \ll 1$. Indeed, the mean-field interaction
is given by the chemical potential $\mu \sim n(t)$, where $n(t)\sim 1/R^2(t)$
stands for a typical density of the expanding cloud scaled by its 
radius $R(t)\approx \sqrt{d^2 + (t/d)^2}$ (in chosen units).   
The interaction-induced additional phase shift can be estimated as  
$\Delta \phi  \approx \int_0^\infty  {\mu dt} \sim \int_0^\infty  {ndt} \approx (d/l_c)^2 \ll 1$, 
where the initial density is taken as $n=1$ in the chosen units.
Hence, under this condition, the density after time $t$ of free expansion becomes
\begin{equation}\label{n}
n({\vec R},t)=\frac{1}{t^3}\left|\int d^3R'
\exp \left( {\frac{{i(\vec R - \vec R')^2 }}{{2t}}} \right)
\Psi _0 (\vec R')\right|^2.
\end{equation}
A numerical factor in (\ref{n}) is set to 1, because it defines only 
overall intensity (not the structure) of the absorption image. 

 When two expanding uniform BECs overlap they form an interference pattern (IP) 
of parallel fringes \cite{Kett1,Kett2}.  
The specific signature of a rotational vortex in the IP is a so-called edge dislocation. 
It was predicted in 
ref.\cite{vipt} and then seen experimentally in ref.\cite{Kett4}. 
Here we will discuss how the JV can be recognized in the IP.
\begin{figure}[h]
\begin{center}
\epsfxsize=8.5cm
\epsfbox{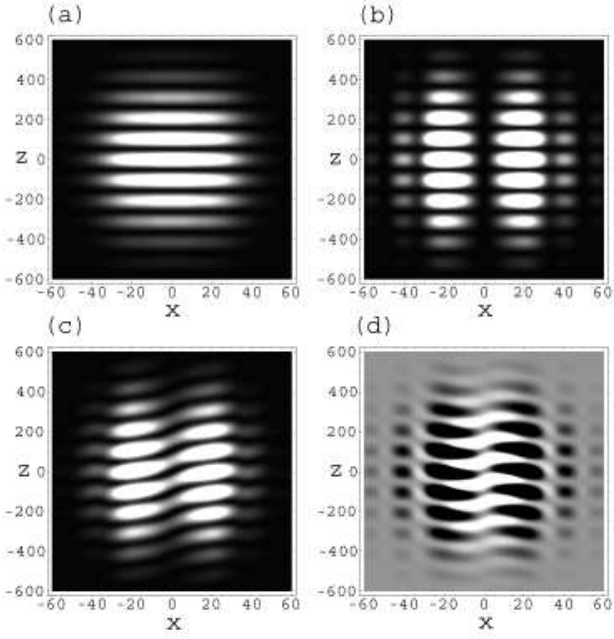}
\caption{
\label{inter}
Interference patterns of two expanded overlapping BEC clouds released from the waveguides
as given by direct numerical integration of (\ref{n}). 
Waveguides initially contained: (a) uniform BECs, (b) two DS aligned at $x=0$, (c) JV 
solution located at $x=0$. The image (d) is the relative intensity between (b) and (c);
 $\nu=0.01$, $t=100$, $d=2^{-3/2}$, $z_0=3$. The unit of x and y is the coherence length $l_c$.
} 
\end{center}
\end{figure}

 For any feature 
with a size $\sim L_s$ comparable with the healing length to become optically resolvable 
it must be enlarged (typically about $10$ times) during the expansion.
We consider the situation when axial expansion of the cloud of length $L$ can be ignored.
This imposes the limitation $\sqrt{t} \ll L$ in the chosen units.
We also ignore quasi-1D thermal fluctuations. In other words, the phase-correlation
length $L_\varphi$ \cite{petrov} is taken larger than the system size $L$. In reality
this is too strong of a requirement --- the discussed feature can be seen when
$L_\varphi \geq L_s$. 

Employing the ansatz (\ref{int_1},\ref{int_2}), the density after time $t$ (such that
$R(t) \gg d$, that is, $t \gg d^2$)
becomes
\begin{eqnarray}\label{dens}
n(x,y,z)&=&\frac{{\rm e}^{-\frac{(y^2 + z^2)d^2}{t^2}}}{t^3}
|\int dx_1{\rm e}^{-ix^2_1/2t}f(x_1)
\nonumber \\
&(&\sin(\frac{zz_0}{t})\cos(\frac{xx_1}{t})\psi''(x_1) + \\
&i&\cos(\frac{zz_0}{t})\sin(\frac{xx_1}{t})\psi'(x_1))|^2
\nonumber
\end{eqnarray}
where the overall factor is dropped, and $\psi'(x)=\sqrt{1+\nu}\tanh(sx)$ and 
$\psi''(x)=Q_0/\cosh(sx)$ are, respectively, the real and imaginary parts of the JV solution
\cite{orig} given by $s=2\sqrt{\nu},\, Q_0=\pm \sqrt{1-3\nu}$.
The signs $\pm$ are due to two possible directions of the current circulation in the JV.    
It is instructive to consider three distinctive cases: i) two uniform
condensates, which can be reproduced from eq.(\ref{dens}) by setting $\psi'=0$ and
$\psi''=1$; ii) two identical DSs at $x=0$, which can be obtained 
by setting $\psi''=0$; and iii) the JV. In the case i), taking the limit $L\to \infty$ and
performing explicit integration, one finds the well known parallel fringes 
$n(x,y,z) \sim \cos^2(zz_0/t)$. In the case ii), there is a zero at $x=0$. Its width
can be estimated from eq.(\ref{dens}) as $\delta x \approx \sqrt{t}$ in the limit $\sqrt{t}\gg 1/s$, that is,
when the DS length $L_s=1/s$ has expanded significantly: $L_s \ll t/L_s$. The actual
density profile can be obtained analytically for $1/s \ll |x| \leq \sqrt{t}$ as
\begin{eqnarray}\label{dens_ds}
n_{DS}(x,y,z)=\frac{4(1+\nu){\rm e}^{-\frac{(y^2 + z^2)d^2}{t^2}}}{t^3}x^2
\cos^2(\frac{zz_0}{t}).
\end{eqnarray}
It features the parallel fringes with the central zero as shown in Fig.3b.

The JV IP can easily be understood by analyzing the vicinity $x=0$. 
Indeed, for   $1/s \ll |x| \leq \sqrt{t}$, the $\tanh(sx)$ function can be replaced
by a step function and  $Q_0/\cosh(sx)$ effectively becomes $ \delta(x)\int dx Q_0/\cosh(sx)=
\pi Q_0 \delta(x)/s $. Thus, the density profile due to the JV becomes
\begin{eqnarray}\label{dens_jv}
n_{JV}(x,y,z)&=&\frac{{\rm e}^{-\frac{(y^2 + z^2)d^2}{t^2}}}{t^3}
(2\sqrt{1+\nu}\cos(\frac{zz_0}{t})\,x  
\nonumber \\
&\pm&\frac{\pi \sqrt{1-3\nu}}{2\sqrt{\nu}}\sin(\frac{zz_0}{t}))^2.
\end{eqnarray}
The profile of the zeros of the density $n=n(x,z,t)$ defines the 
feature specific for the JV.
In the DS case, zeros belong to the set of mutually orthogonal lines 
$x=0$ and $z=\pi(n + 1/2)t/z_0$, with $n$ integer.
In the JV case, represented by eq.(\ref{dens_jv}),
the lines of zeros do not cross any more and obey the condition 
\begin{equation}\label{zerJV}
x =\pm \frac{\pi }
{4}\sqrt {\frac{{1 - 3\nu }}
{{\nu (1 + \nu )}}} \tan \left( {\frac{{sz}}
{t}} \right)
\end{equation}
The inclined tangential feature seen on Fig.3c  is a 
consequence of the smooth relative phase change from $0$ to $2\pi$ in the JV. 
It is also worth noting that the JV circulations in different directions produce tangential slope of different sign in the IP. 

On  
Fig.(\ref{inter}) we have plotted column densities (integrated over $y$) at $t=100$ 
($t=50ms$ in usual units, which is a typical experimental expansion 
time after which the absorption image is taken).
It should be noted that the above presented IPs correspond to
the case when the separation between the waveguides $z_0$ is significantly
larger than the  transverse extension $d$. Obviously, in this situation
the tunneling $\nu$ is essentially zero. If one tries to decrease $z_0/d$, 
the visibility of the fringes worsens due to the exponential factor in (\ref{dens})
so that for $z_0/d \sim 1$ just one central
fringe is seen. 
Thus, in order to achieve a good resolution, the clouds 
should be quickly separated from each other and, then, immediately released. 
Under these conditions, the JV solution formed at a closed proximity between 
the waveguides will have no time to be distorted by the inter-particle interactions
after the tunneling is cut off. Obviously, the duration of the waveguides'
separation from some distance $z_0 \approx d$ (when the tunneling is finite) 
to, e.g., $z_0 \approx 10 d$ (when the tunneling is, practically, zero) is limited from below
by the inverse frequency $ \approx d^2$ of the radial confinement. From above, it is limited by the
axial response time $1/\mu \approx l_c^2$. This requirement can safely be satisfied if $(d/l_c)^2\ll 1$,that is,
in the quasi-1D regime.

The above discussion has been limited to destructive imaging of the JV. Very recently, a non-destructive method 
has been employed \cite{lscsi}, \cite{weakln} to continuously sample the relative phase of two spatially separated BECs. 
Tilted fringes in the interference pattern of the outcoupled matter waves have been seen as an evidence of axial gradients of the relative phase \cite{weakln}. This method can also be very usefull for detecting the Josephson vortex and its conversion into the dark soliton and vice versa. The outcoupling pulse produces recoil atoms characterized by the wavefunction which, in the co-moving frame, is a replica (apart from the numerical factor) of the wavefunction of the confined atoms. Thus, the interference pattern produced by the outcoupled clouds will be identical to the one discussed above. 

\section{Conclusion}

The atomic Bose Josephson vortex can be created by the phase imprinting 
technique and detected due to its particular feature in the column density
by absorption imaging performed after some ballistic expansion. 
The Josephson vortex can be controllably displaced by
imposing tunneling current (created by disbalance of chemical potentials) between the waveguides. 
In quasi-1D, motion of an atomic Josephson vortex is strongly coupled
to the current circulation through the phase-slip effect. 
This leads to a destruction of the circulation for the vortex speeds above a certain value
determined by the Josephson coupling. In contrast to the standard approach to Josephson vortices in superconductors 
within the Sine-Gordon formalism, a description of the coupling between the center of mass motion and the circulation necessarily involves both density and phase variations.

\section{acknowledgement}
This work
is supported by the NSF grant PHY-0426814 and by
the PSC-CUNY grant 66556-0035.

\end{document}